\documentstyle[rmaaconf]{article}

\begin{document}

\title{STARBURST ANATOMY:  STELLAR AND NEBULAR PROPERTIES \\ OF 
NEARBY GIANT HII REGIONS}

\author{William H. Waller\altaffilmark{1}}
\altaffiltext{1}{Hughes STX Corporation, NASA Goddard Space Flight Center, 
Laboratory for Astronomy and Solar Physics, Code 681, Greenbelt, MD  20771, 
U.S.A.; E-mail (waller@stars.gsfc.nasa.gov)}

\centerline {To appear in {\it Revista Mexicana de Astronomia y Astrofisica}}

\begin{resumen}
Resolution of nearby giant HII regions into their stellar and nebular 
constituents provides fundamental insights for interpreting more distant and 
powerful starburst activity.  The following summarizes recent advances in our 
understanding of the stellar populations and nebular 
energetics associated with giant HII regions.
\end{resumen}

\begin{abstract}
Resolution of nearby giant HII regions into 
their stellar and nebular constituents provides fundamental insights for 
interpreting more distant and powerful starburst activity.  
Photometry and spectroscopy of 
ionizing clusters in the Galaxy, LMC, and SMC reveal no significant relation 
between metallicity and the slope of the power-law initial stellar mass 
function (IMF).  HST/WFPC2 photometry of 3 giant HII regions in M33 also does 
not show any consistent trend involving metal abundance and IMF slope --- 
contrary to predictions based on emission-line ratios.  The upper stellar mass 
limit appears to be constrained more by cluster age than by anything else.

The ionizing luminosities from some of the resolved stellar populations are 
insufficient to account for the ionization rates inferred from measurements of 
the composite (integrated) H$\alpha$ emission.  Absorption of stellar EUV 
emission by nebular dust grains would only amplify these photoionizing 
shortfalls.  Leakage 
of ionizing photons from the HII regions would further exacerbate the 
situation.  What then is providing the additional ionization?  Prospects for 
higher stellar EUV luminosities and/or alternative sources of nebular 
ionization (e.g. shocks) are evaluated with this question in mind.
\end{abstract}

%\keywords{\bf Galaxies: Starburst, Star Clusters --- ISM:  Dust, Extinction, 
%HII 
%Regions --- Stars:  Early-Type, Luminosity Function, Mass Function --- 
%Ultraviolet:  Stars}

\section{Key Questions}
  
A fundamental question regarding starburst activity in galaxies is ``What are 
starbursts made of?''  More specifically, (1) how are the ionizing stellar 
populations clustered, (2) how does the stellar initial mass function depend on 
starburst intensity, metallicity, and other environmental factors, and 
(3) what balance of radiative and 
mechanical processes powers the starburst nebulae?  A full answer to these 
questions requires the ability to 
resolve the stellar and nebular constituents of the starbursts.  Fortunately, 
this can be done for giant HII regions (GHRs) 
in the Galaxy and in other galaxies of 
the 
Local Group.  The following discussion of stellar and nebular properties is 
intended to complement the related reviews by C. 
Leitherer, A. Moffat, and H. Zinnecker in these Proceedings.

\section{Starburst Constituents vs. Available Resolution}

The ability to resolve stellar and nebular structure in starbursts 
depends on the distance to the starbursting source, the source's 
degree of central concentration, and the angular resolving power that is 
available.  
The Orion nebula provides a handy benchmark 
for ascertaining the degree of resolution attainable with each source.
Orion's central ``Trapezium'' cluster 
of hot stars spans approximately 15$''$ (0.04 pc), while its most prominent 
nebular feature --- the ionization front to the SE --- 
is roughly 10-times larger in 
extent ($~\sim$0.4 pc).  These stellar and nebular 
features can be resolved from the ground out to distances of 
$\sim$10 kpc \& 100 kpc respectively.  Imaging with the HST extends the ``fully 
resolved'' stellar and nebular structures to distances of 
100 kpc \& 1 Mpc respectively.  At the distance of M33 (0.84 Mpc), 
the dense clustering 
seen in Orion, NGC 3603, and 30 Doradus would remain partially resolved.  And 
beyond 1 Mpc distance, 
the probability of mistaking compact clusters as point sources becomes 
problematic.

However, dense cluster formation is not universal to nearby starburst activity 
--- despite the impressive numbers of luminous and compact clusters that 
have been found recently in starburst environments (Ho, these Proceedings).
Indeed, the degree of clustering is found to vary 
greatly among nearby starbursts 
--- 
from compact ``super star clusters'' such as the R136 core in 30 
Doradus, to the sprawling stellar populations evident in the giant HII regions 
of M33.  For example, R136 
contains about 
17,000 stars 
(M $\ge$ 4 M$_{\odot}$) within a diameter of 10 pc, or roughly 2.3 times the 
number of stars 
that is scattered across the central 100 
pc of NGC 595 --- 
the second most massive GHR in M33 (Malumuth, Waller, \& Parker 1996).  
HST/WFPC2 imaging of NGC 604, the most massive GHR in M33, reveals a similarly 
loose 
association of $\sim$35,000 OB stars sprawling over a 100 pc diameter field 
(Hunter et al. 1995). 
What these loosely clustered starbursts 
lack in central concentration is compensated through their larger 
sizes.  Moreover, 
their less crowded populations enable stellar photometry and 
spectroscopy to be conducted in galaxies well 
beyond the ``fully resolved'' distance 
of 100 kpc.

The nebular structure of giant HII regions is exemplified by the exquisitely 
complex views of 30 Doradus and NGC 604 as obtained at H$\alpha$ with 
groundbased telescopes and the HST/WFPC2 (cf. Cheng et al. 1992; 
Hunter et al. 1995).
In both GHRs, the embedded OB clusters have evacuated inner cavities which, in 
turn,  
enable them to illuminate 
myriad bright rims, loops, and bubbles in the surrounding gas.  
Coherent structures 
are evident on scales of $\sim$100 pc down to the resolution limit of $\le$1 
pc.  The dynamical origins and current energetics of these nebular structures
remain uncertain (see 
Section 4).

\section{Stellar Populations --- A Remarkable Consensus Emerges}

Composite indices of the stellar populations powering starburst activity 
include the broadband visible and UV colors, absorption-line strengths and 
ratios, hydrogen-line luminosities and equivalent widths, emission-line ratios, 
and far-infrared ``excesses'' relative to the radio Bremsstrahlung emission.  
Such composite diagnostics have led to claims of environmentally-sensitive 
IMFs.  For example, studies of emission-line ratios in GHRs and HII galaxies 
indicate IMFs that are biased towards hotter, higher-mass stars in regions of 
lower metal abundance (Campbell 1987; Vilchez \& Pagel 1988).  Other studies 
involving the H$\alpha$ equivalent width as a tracer of the high-mass IMF 
have found no metallicity dependence, but rather a sensitivity to dynamical 
factors such as shearing and tidal disruption in the disks of galaxies (Waller 
1990; Waller \& Hodge 1991).  Even more provocative are 
multi-diagnostic studies which infer a high-mass bias in regions of especially 
intense 
starburst activity (Gehrz, Sramek, \& Weedman 1983; Rieke et al. 1993).  

Such claims of environmentally-sensitive IMFs depend on the assumption that one 
has correctly interpreted the various composite spectral indices.  To evaluate 
the composite indices in terms of the actual stellar populations, it is first 
necessary to {\it resolve} and photometrically characterize the prominent 
members of the population.  Spectroscopic follow-up of the hottest and 
brightest members is then necessary to constrain the temperatures, ionizing 
luminosities and massies of the stars most responsible for powering the 
composite spectral indices.  In the past decade, considerable efforts have been 
made to characterize the stellar populations powering GHRs in the Galaxy and 
in other 
galaxies of the Local Group (see reviews by Leitherer, Moffat, and Zinnecker in 
these Proceedings).  
{\it What is remarkable is the degree of consensus that has 
been reached:  In all nearby GHRs studied so far, the stellar 
IMF appears similar to 
that found in smaller clusters near the Solar 
neighborhood}.

Photometry and spectroscopy of ionizing clusters in the Galaxy, LMC, and SMC 
do not reveal any significant relation between metallicity and the 
slope ($\Gamma$) of 
the power-law IMF.
Multi-band HST/WFPC2 photometry of 3 GHRs in M33 also does 
not show any consistent trend involving metal abundance and IMF slope 
(Hunter et al. 1995; Malumuth, Waller, \& Parker 1996; Parker et al. 1996).
Over a one-dex range of O/H abundances and a two-dex range of 
H$\alpha$ luminosities, the IMF slopes of 11 GHRs average 
to $<\Gamma> = -1.3 \pm 0.2$ (Waller 1996).
The upper-mass limit appears to be constrained more by cluster age 
than by anything else.  And the proportion of lower-mass stars 
appears to be consistent with the derived IMF slopes 
down 
to detection limits of 2 M$_{\odot}$ in 30 Doradus and 0.2 M$_{\odot}$ in the 
Orion nebula (Zinnecker, these Proceedings).  
The results from these resolved studies strongly 
differ with those based on the 
analysis 
of composite emission-line ratios (Campbell 1987; Vilchez \& Pagel 1988; 
Vilchez et al. 1988) and other composite spectral diagnostics (Gehrz, Sramek, 
\& Weedman 1983; Rieke et al. 
1993).

\section{Nebular Energetics --- Balancing the Power}

From the Orion nebula to the GHRs in M33, the ionizing luminosities of the 
resolved clusters barely account for the ionization rates that are 
inferred from 
measurements of the composite (integrated) H$\alpha$ emission.  In some GHRs 
significant {\it shortfalls of ionizing photons are measured relative to 
the gaseous ionization} (Malumuth, Waller \& Parker 1996; Parker et al. 1996; 
Waller 1996).  
Absorption of stellar EUV emission by nebular dust grains 
could drastically amplify the photoionizing shortfalls (Aannestad 1989).  
Leakage of ionizing photons from the HII 
regions could further exacerbate the situation (Gonzalez-Delgado et al., 
Wall, these Proceedings).  
What then is providing the 
additional ionization?

One possibility is that the ionizing luminosities of O-type stars have been 
underestimated.  EUV fluxes of O-type stars 
have yet to be measured, and hence must be estimated by extrapolation of models 
based on UV and optical spectra.  Currently, the only EUV measurement of an 
early-type star is of the B2 II star Epsilon Canis Majoris (Cassinelli et al. 
1995).  The resulting Lyman continuum luminosity of this star exceeds that 
predicted from optical and UV spectroscopy by a factor of 30!

Meanwhile, theoretical models of ionizing clusters are rapidly evolving, as new 
methods are developed for treating the EUV opacity in the hot and windy stellar 
atmospheres (Garcia-Vargas et al. 1995; 
Najarro et al. 1996).  Some of these models predict strong declines in the EUV 
luminosities with increasing metallicity, possibly explaining the downturns in 
H$\alpha$ equivalent widths and [OIII]/[OII] line ratios that are observed 
at higher metallicity (Waller 1990; Campbell 1987; Vilchez \& Pagel 1988; 
Vilchez et al. 1988; Shields 1990).  
Other models which include 
powerful stellar winds are able to reduce the EUV opacities and thereby 
increase the ionizing luminosities well-above those predicted by the less windy 
models (Najarro et al. 1996).  

Another possible explanation for the apparent shortfall of ionizing luminosity 
from the resolved clusters is that shock waves are contributing significantly 
to the nebular ionization.  Sufficiently strong shocks can arise in the 
bubbles blown by high-mass stars, in the   
supersonic turbulence generated by outgassing virialized low-mass stars, and in 
the filamentary shells that are driven by supernova explosions (cf. Losinskaya 
1992; Tenorio-Tagle et al. 1996).  All of these energetics are present in giant 
HII regions, whose contents can include 
10$^5$ -- 10$^6$ of low-mass stars with supersonic velocity dispersions, 
hundreds of O stars, tens of WR stars exhibiting superwinds, 
and the potential for tens of 
supernovae every Myr (Malumuth, Waller, \& Parker 1996).  

Constraining the various sources, 
sinks, and pathways of power in nearby GHRs remains an important challenge.  
Towards this end, several investigations of the resolved stellar and nebular 
properties in GHRs 
are in progress. 
By addressing {\it both} the radiative and mechanical energetics, 
these 
studies will provide 
fundamental insights for 
interpreting the unresolved activity in more distant and 
powerful starbursts.

\end{document}